\documentclass[twoside,11pt]{article}

\usepackage[letterpaper, margin=1in]{geometry}
\usepackage{natbib}
\usepackage{microtype}
\usepackage{graphicx}
\usepackage{wrapfig}
\usepackage{mdframed}
\usepackage{subcaption}
\usepackage{hyperref}
\usepackage{xcolor}
\usepackage{amsmath}
\usepackage{amssymb}
\usepackage{mathtools}
\usepackage{enumitem}
\usepackage{csquotes}
\usepackage[normalem]{ulem}
\usepackage{booktabs}
\usepackage{bm}
\usepackage{algorithm}
\usepackage{algcompatible}
\usepackage{algpseudocode}
\DeclareGraphicsExtensions{.pdf}

\setlist[itemize]{labelindent=16pt}
\setlist[enumerate]{labelindent=16pt}
\setlist[description]{font=\normalfont\space,labelindent=16pt,leftmargin=48pt}

\DeclareMathOperator*{\argmin}{arg\,min}

\begin{document}

\title{A predictive processing model of perception and action for self-other distinction} 
\author{
	\normalsize
	\textbf{Sebastian ~Kahl}\footnote{Corresponding author: \texttt{skahl@uni-bielefeld.de}},
    \textbf{Stefan ~Kopp}
    \\
    \normalsize
    Social Cognitive Systems Group, CITEC, Bielefeld University, Bielefeld, Germany
}
\date{\vspace{-5ex}}
\maketitle

\begin{abstract}
During interaction with others, we perceive and produce social actions in close temporal distance or even simultaneously. It has been argued that the motor system is involved in perception and action, playing a fundamental role in the handling of actions produced by oneself and by others. But how does it distinguish in this processing between self and other, thus contributing to self-other distinction? In this paper we propose a hierarchical model of sensorimotor coordination based on principles of perception-action coupling and predictive processing in which self-other distinction arises during action and perception. For this we draw on mechanisms assumed for the integration of cues for a sense of agency, i.e., the sense that an action is self-generated. We report results from simulations of different scenarios, showing that the model is not only able to minimize free energy during perception and action, but also showing that the model can correctly attribute sense of agency to own actions.
\end{abstract}

\section{Introduction}
% whats to problematic about social interaction?
In everyday social interaction we constantly try to deduce and predict the underlying intentions behind others' social actions, like facial expressions, speech, gestures, or body posture. This is no easy problem and the underlying cognitive mechanisms and neural processes even have been dubbed the ,,dark matter'' of social neuroscience \citep{Przyrembel:2012dua}. 
Generally, action recognition is assumed to rest upon principles of prediction-based processing \citep{Clark:2013jo}, where predictions about expected sensory stimuli are continuously formed and evaluated against incoming sensory input to inform further processing. Such a predictive processing does not only inform our perception of actions of others, but also our action production in which we constantly predict the sensory consequences of our own actions and correct them in case of deviations. Both of these processes are assumed to be supported by the structure of the human sensorimotor system that is characterised by perception-action coupling \citep{Prinz:1997uu} and common coding of the underlying representations. 

It seems natural to assume that these general mechanisms of prediction-based action processing underly also the perception and generation of social behavior when we interact with others. However, in dynamic social interaction, perception and action often need to be at work simultaneously and for both, actions of self and other. If the motor system is to be involved in both processes, this simultaneous processing and attribution of actions to oneself and the interaction partner must be maintained during social interaction without confusion.

As of yet, it is not clear how self-other distinction is reflected in, or possibly even constituted by the sensorimotor system. What role does it play in distinguishing social actions of the self and the other? What are the underlying prediction-based processes? And how do they interact with higher-level cognitive processes like mentalizing to solve the social differentiation problem? Evidence from cognitive neuroscience suggests that the motor systems may be involved differentiately in processing self-action and other-action, indicating a role in social cognition \citep{SchutzBosbach:2006bi} and the selective attribution of beliefs, desires and intentions during the dynamic process of intersubjective sense-making. We contribute a computational modeling perspective. In previous work we devised a model of the interplay of mentalizing and prediction-based mirroring during social interaction \citep{Kahl:2015uy}. In that work two virtual agents interacted in a communication game, each of which equipped with models of a mirroring system and a mentalizing system, respectively. This demonstrated how mentalizing -- even with minimal abilities to account for beliefs, desires and intentions -- affords interactive grounding and makes communication more robust and efficient. However, both agents took separate and successive turns such that their mirroring systems worked either for perception or production of social actions (albeit with activations being carried on).

In this paper, we present an extended Bayesian model of a sensorimotor system based on a prediction-based processing hierarchy, called Hierarchical Predictive Belief Update (HPBU). Our aim is to explore how such a prediction-based sensorimotor system can be able to differentiate by itself actions of its own (predicted for production) from potential actions of the interaction partner (predicted for perception), thus contributing to self-other distinction in social interaction. 

For our computational model of sensorimotor processes and the model of sense of agency we rely on assumptions from the predictive processing framework. Specifically, the model will rely on principles of active inference and free energy minimization \citep{Friston:2010ez} based on assumptions in accord with the predictive brain hypotheses \citep{Clark:2013jo} and hierarchical predictive coding \citep{Rao:1999in}. From that follows a hierarchical organization of ever more abstract predictive representations. These representations are also generative processes, which together form a hierarchical generative model which maps from (hidden) causes in the world to their perceived (sensory) consequences. The tight coupling between action and perception in active inference means, that following prediction errors, either the model hypotheses have to be updated or action in the world is necessary to make future sensory evidence meet the model predictions. 
Free energy is merely the term for the negative log model evidence of a perceived event given the model prediction, i.e., the prediction error which is to be minimized. Technically, variational free energy is an upper bound on Bayesian model evidence, such that minimizing free energy corresponds to minimizing (precision-weighted) prediction error – or, equivalently, maximizing model evidence or marginal likelihood.

Also, we will identify, integrate and then test mechanisms and processes in the SoA (sense of agency) literature that are compatible with a predictive processing view and that have reliably been identified to contribute to sense of agency. Our goal here is to present a conceptual computational model of the sense of agency which is functionally embedded in a hierarchical predictive processing model for action production and perception. The model of sense of agency itself integrates different aspects important for sensorimotor processing and motor control. At the core of our argument for the presented modeling approach is the assumption that we strongly rely on the predictability of our own body to be able to not only identify our own hands or arms, but also to differentiate between our own and other's actions through the information gathered especially on the unpredictability of others. It is that unpredictability of their actions and it's timing that can help to differentiate. The functional simulation at the end of our paper helps to evaluate whether the identified mechanisms implemented in the model are sufficient to correctly infer own actions, given altered feedback to its action production. 

We start with introducing HPBU and how it forms, tests, and corrects so-called motor beliefs. Then we discuss how this model can be extended with a mechanism for differentiating between actions produced by oneself from actions produced by a potential interlocutor in social situations. The mechanism includes the ability to flexibly integrate predictive and postdictive cues to form a sense of agency (SoA). Finally, we present results from simulations that test the model's ability to infer SoA for its own actions in different test scenarios. We then discuss our results with the literature concerning the mechanisms underlying SoA and discuss the mechanism's implications for the process of attributing beliefs during communication.
% ------------------------------------------------------------------------------------
\section{Hierarchical Predictive Belief Update}
We adopt a Bayesian approach to computationally model the human sensorimotor system at a functional level. The model, called Hierarchical Predictive Belief Update (HPBU) realizes an active inference and free energy minimization. Doing so it is able to form, test and correct so-called motor beliefs in perception and production of actions. Based on our previous work \citep{Sadeghipour:2010gy} and other attempts to model the sensorimotor system \citep{Wolpert:2003gm}, we chose to make use of a hierarchical representation of increasing abstractions over representations of movement. Each level contains a generative model that infers probabilities to perceive (and produce) these variants of abstractions over actions in the form of discrete probability distributions about discrete representations that can be influenced both bottom-up, in the form of evidence for its last prediction from the next lower level, and top-down in the form of a prediction by the next higher level. Following the assumption that the main flow of information is top-down and that motor control is also just top-down sensory prediction or "active inference", all levels receive their next higher level's prediction and evaluate it for their own bottom-up prediction in the next time step. The distinction to previous models of the sensorimotor system is that in active inference we solely rely on each level's generative process to map from (hidden) causes to their (sensory) consequences. Without separate inverse models the generative process itself is inverted to predict the next steps in the next lower level and thus, explain away or suppress prediction errors. In the lowest level of the hierarchy this suppression can take the form of triggering the production of actions and change the environment as to minimize prediction errors.

The representations in our hierarchy code for both, the perception and production of an action and in that follow the common-coding theory of perception-action coupling \citep{Prinz:1997uu}, a defining characteristic for representations in the mirror neuron system \citep{GALLESE:1996vm}. The human mirror neuron system has indeed also been attributed to have a hierarchical organization, which is distributed across interconnected brain areas \citep{Grafton:2007jf} and similarly, predictive coding and active inference have repeatedly been linked to the mechanisms underlying the function of the mirror neuron system \citep{Kilner:2007jq,Friston:2010ez}.

% mention previous predictive coding work on the attribution of agency
Also, in the predictive coding and active inference literature the attribution of agency was attributed to rely on mechanisms central also to the model presented in this paper, i.e., the correct prediction of the consequences of producing actions of handwriting \citep{Friston:2011cua}. But other than the mechanisms we will later go into, they rely heavily on the proprioceptive information which is missing when perceiving other's actions in contrast to actions performed by oneself. We will argue that there is sufficient information already available in the visual information only.
\begin{figure}[t]
\begin{center}
\includegraphics[width=10cm]{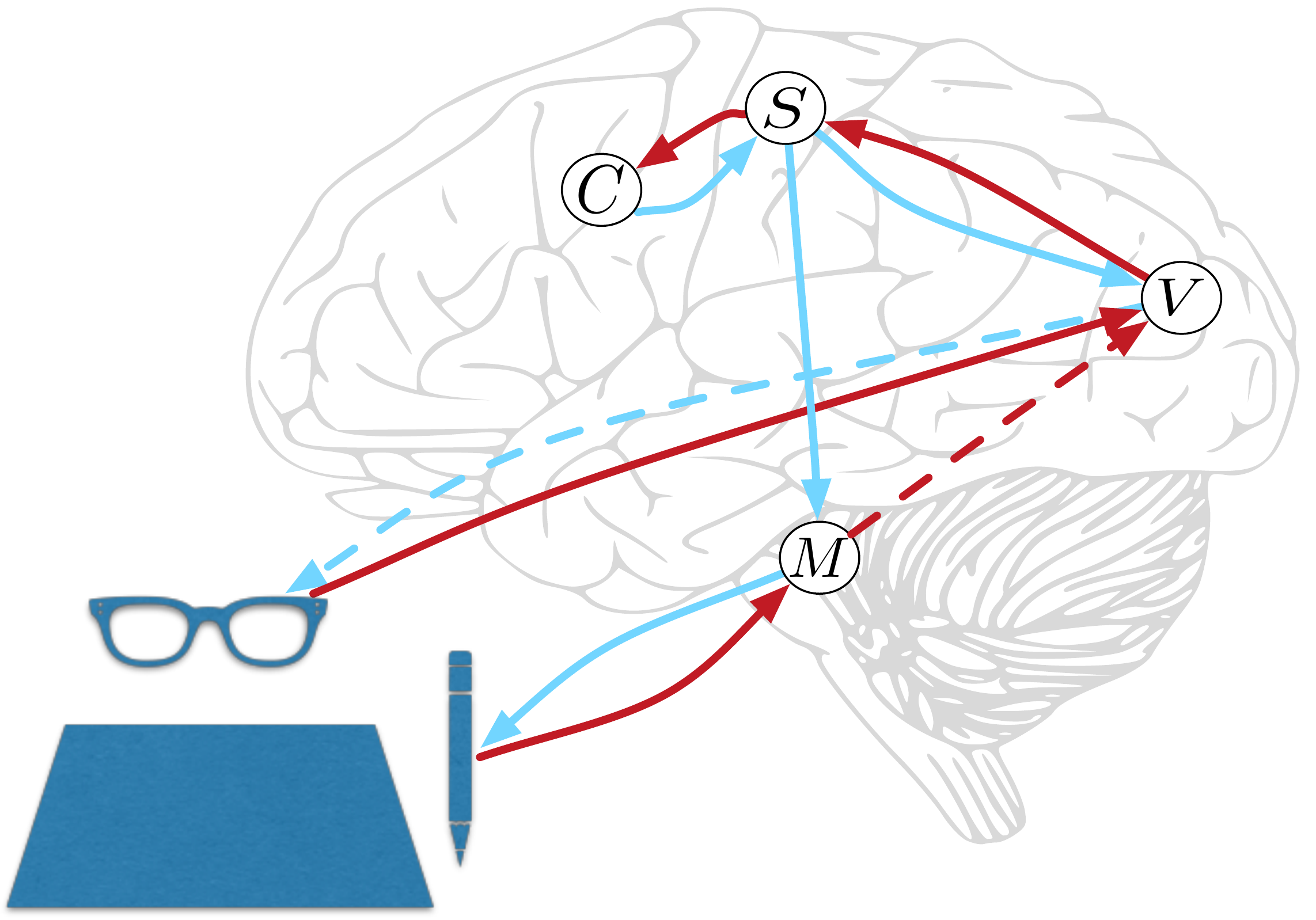}
\end{center}
\textbf{\refstepcounter{figure}\label{fig:model_hierarchy_figure} Figure \arabic{figure}.}{Hierarchical Predictive Belief Update, based on predictive processing and perception-action coupling. Predictions are sent top-down and compared with sensory (bottom-up) evidence to drive belief updates within the hierarchy. We have loose associations with the displayed cortical and subcortial structures. The connections shown will be explained in great detail below. At the top, in the \emph{Schema} level (\emph{C}) we represent abstract clusters of action sequences grouped by similarity. Below that, the \emph{Sequence} level (\emph{S}) represents sequences of motor acts. The lowest levels in the model hierarchy allow for action production, and proprioceptive feedback in \emph{Motor Control} level (\emph{M}) and visual input and action feedback in the \emph{Vision} level (\emph{V}). Red and blue lines represent \emph{bottom-up} and \emph{top-down} information propagation, respectively. The blue dotted line from \emph{V} represents a visual prediction without any effect on the world, while the blue line from \emph{M} has a causal effect. The red dotted line from \emph{M} represents a long range connection, further explained below.} 
\end{figure}
HPBU can be described as a three-level hierarchy of motor beliefs as abstractions over primitive motor acts (see Figure~\ref{fig:model_hierarchy_figure}). At the top, in the \emph{Schema} level (\emph{C}) we represent abstract clusters of action sequences grouped by similarity. Below that, the \emph{Sequence} level (\emph{S}) represents sequences of motor acts. 
Levels in our hierarchy are loosely associated with the following (sub-)cortical structures: The Vision level corresponds to early level V5 in the visual cortex for its direction selectivity in the perception of motion, while the Motor Control level corresponds to the reflex arcs embedded in the tight coordination of basal ganglia, spinal coord and cerebellum. The Sequence and Schema levels are loosely associated with primary motor and premotor areas that code for representing action sequences.
We assume these representations to be the basis for active inference for both action perception and production. Further, we assume the representations to be multimodal, i.e., combining visual, motor and proprioceptive aspects of action, if available. Consequently, they are used as more or less high-level or visuomotor representations of action and their outcomes. This is based on converging evidence for the multimodal nature of representations that can be found in somatosensory, primary motor and premotor areas of the human brain, which can code for both visual and proprioceptive information \citep{wise_boussaoud97,Fogassi:2005cg,Pipereit:2006ig,Zhang:2009vd,Gentile:2015ia}. The lowest levels in the model hierarchy allow for action production and visual and proprioceptive input and feedback in the form of two separate models; the \emph{Vision} level (\emph{V}) and \emph{Motor Control} level (\emph{M}) that will be further described below.
% ------------------------------------------------------------------------------------
\subsection{Motor Coordination}
In motor control two problems have to be solved. First, how to learn action sequences in order to reach a goal, i.e., a mapping from an extrinsic coordinate frame (describing the action perception) into an intrinsic frame (describing the muscle movement) and second, how to activate the appropriate muscles to reach a goal, i.e., from an intrinsic frame to an extrinsic frame to produce the desired movement. What makes this problem hard is that there are many possibilities how the intrinsic frame could produce the extrinsic frame.

To solve the problem of action production towards a goal often the planning of an optimal trajectory was assumed using forward models trying to find an optimal sequence of muscle activations leading to the reached goal before the action even starts \citep{Kawato:1999gn}. \cite{Todorov:2002bf} brought up a different solution trying to explain the high variability in the detailed movements that occured even in repeated actions. They proposed a control strategy that allowed for variability in redundant task dimensions during the action production. That is, during action production feedback is used to only correct variability that interferes with reaching the goal. The use of feedback was often discussed in this context, famously the MOSAIC model by \cite{Wolpert:2003gm} proposed how using a comparison of produced action and its sensory feedback could stabilize and guide action. What counts as feedback in this context are the visually perceived positions of the joints controlled by the muscles and the proprioceptive feedback by the muscle spindles surrounding the muscles. 

An important distinction on such models of optimal control is highlighted by \cite{Friston:2011dw} in that the forward models in motor control are not the generative models used in perceptual inference and hence should not be conflated. He argues that one could get rid of the forward models for action by replacing the control problem with an inference problem over motor reflex arcs and in that simplifying optimal control to be active inference. In active inference the extrinsic frame can be utilized as action-production tools, to circumvent the need for detailed programming of motor commands. 

We follow this argumentation. First, we transform the perceived action into a gaze- or vision-centered oculomotor frame of reference that has been shown to also code the visual targets for reaching and other actions \citep{Ambrosini:2012ku,Engel:2002hw,Russo:1996dl}. Using those we are also able to circumvent the need for detailed programming of motor commands. Instead, the oculocentric frame of reference can guide the actions or parts of an action.

The lowest levels of the hierarchy represent two aspects in active inference that are necessary for motor coordination.
The \emph{Vision} level receives continuous coordinates of a writing trajectory at discrete points in time, which is perceived in the form of a discrete probability distribution over a discrete set of writing angles at each point in time. Following the narrative by \cite{Zacks:2007bk,Gumbsch:2017ub} on event segmentation using surprise as a separator, we identify surprising deviations of the current writing angle given the writing angles of the past. In the context of free energy minimization, a sudden increase in the amplitude of prediction errors – induced by a surprising event – is reflected by free energy increases in the \emph{Vision} level. The surprising stroke consists of the writing angle and its length, which are both transformed into the oculocentric reference frame, i.e., into relative polar coordinates with the last surprising stroke coordinates at its center. This information is send to the \emph{Sequence} level, together with the time passed since the last surprising stroke. 
The \emph{Sequence} level stores these sequences of surprising events in the oculocentric reference frame, which can also be used for generation. Following the argument by \cite{Friston:2011dw} we circumvent the need for detailed programming of motor commands by utilizing the surprising events consisting of relative polar coordinates as action targets, which are send to the \emph{Motor control} level. There, a reflex arc in the form of a damped spring system (inspired by \cite{Ijspeert:2013bt}) will realize the motion towards the action target following simple equations of motion with the spring's point of equilibrium at the relative polar coordinate of the action target. This implementation of active inference is formally related to the equilibrium point hypothesis \citep{Feldman:1995ia}. In other words, the top-down or descending predictions of the proprioceptive consequences of movement are regarded as setting and equilibrium or set point to which the motor plant converges, via the engagement of motor reflexes. This will later be explained in more detail.

An important aspect in motor coordination model can be seen in Figure~\ref{fig:model_hierarchy_figure}, where the Motor Control level has no direct feedback connection to the Sequence level. This is for three reasons. First, we would like to see if sequential motor coordination as well as the inference of a sense of agency are possible with visual feedback only. Second, other than \cite{Friston:2011cua} who rely heavily on proprioceptive information we want to allow for visual input to drive motor coordination and a sense of agency. Instead, we close the motor coordination loop using a direct long range connection that is used by the Motor Control level to inform the Vision level when it is done coordinating actions to reach a subgoal (see red dotted line in Figure~\ref{fig:model_hierarchy_figure}). Vision level will then check if visual information can confirm the movement, then sending the information to the Sequence level, closing the motor coordination loop. Third, making the model's sequence coordination independent from direct proprioceptive feedback allows for future developments that can associate actions in the world with intended effects that not directly influence the motor system, e.g., switching on a light or influencing another agent's beliefs.

During action perception, we further assume that the correspondence problem is solved in the sense that an observed action by another agent is mapped into one's own body-centered frame of reference. That is, we feed the perceived action trajectory directly and bottom-up into HPBU. The next section will describe the active inference and free energy minimization in the model hierarchy.
% ------------------------------------------------------------------------------------
\section{Model update details}
The HPBU model is defined as a hierarchical generative model which learns to predict and explain away prediction errors and in this sense minimize the free energy. This section will briefly introduce our free energy minimization strategy and the generative model updates, which we have implemented and extended to encompass and allow the distinction of self and other in the context of action production. 

In the hierarchical generative model, each level maps the internal discrete state space from one level in the hierarchy to the domain of it's next lower level. Each level contains a discrete probability distribution about that level's discrete states. The difference between continuous and discrete states in the context of active inference is well discussed in \cite{Friston:2017de}. All levels of the hierarchy are updated sequentially starting at the top in Schema level \emph{C}, i.e., they are updated in sequence from it's next higher and next lower levels, learning to represent and produce the states in the next lower level and the environment. At the top, the Schema level \emph{C} represents clusters of similar representations of its next lower level, which is the Sequence level \emph{S}. \emph{S} represents sequences of representations of Vision level \emph{V} that occured over time. Also, representations that can map to \emph{V} are compatible with representations in Motor Control level \emph{M}. In each level posteriors are updated bottom-up and top-down and free energy is calculated with respect to prior and posterior distributions as described in the following. These mappings can be understood as a generative processes, where one level predicts the states of the level below. In the Motor Control level \emph{M} this mapping results in action production.
% ------------------------------------------------------------------------------------
\subsection{Free energy minimization}
Free energy describes the negative log model evidence of a generative model that tries to explain hidden states, i.e., the environment. Evidence corresponds to probabilities of data from the environment, given the model at hand.

\begin{figure}[t]
\begin{center}
\includegraphics[width=14cm]{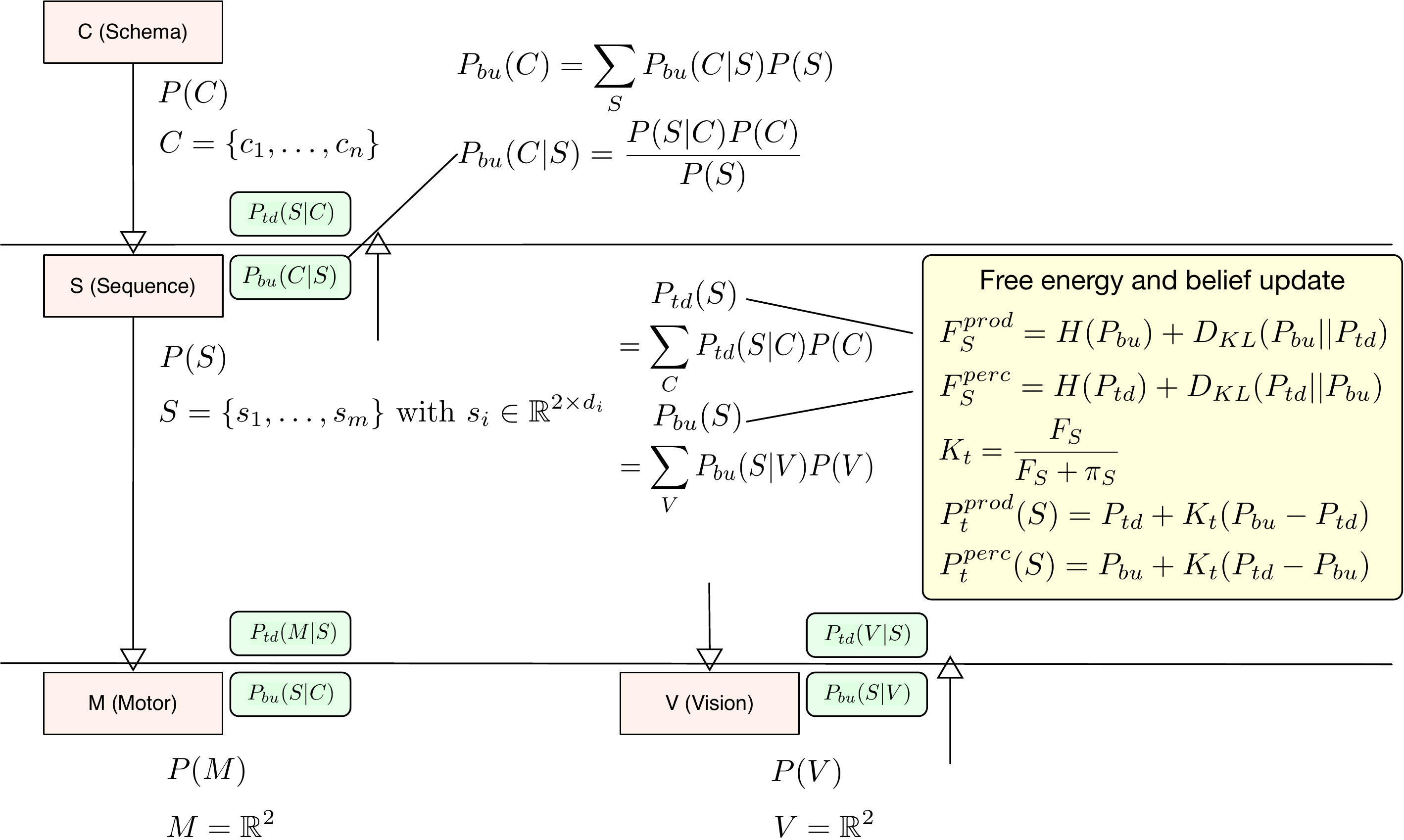}
\end{center}
\textbf{\refstepcounter{figure}\label{fig:model_update_summary} Figure \arabic{figure}.}{To exemplify the belief update scheme in HPBU here the updating of posteriors and free energy within Sequence level \emph{S} are shown. Within \emph{S} the top-down and bottom-up posteriors are calculated from posteriors of \emph{C} and \emph{V}. That information is used to update free energy either for production (\emph{prod}) or for perception (\emph{perc}). The updated free energy and precision ($\pi$) are used to calculate the Kalman gain necessary for calculating the level posterior \emph{P(S)} using a Kalman filter, again either for production or perception. The green boxes show the calculated likelihoods necessary for updating within \emph{S} or in the levels above and below ($P_{bu}(C|S)$ is one example). For more details please refer to the Supplementary Materials.} 
\end{figure}

The free energy in system $\mathbb{X}$ is expressed as the sum of surprise and a cross entropy of two states (a posterior \emph{P}, and a prior \emph{Q} before evidence has arrived). When free energy is minimized the cross entropy becomes zero, thus leaving free energy to be just surprise (entropy with regard to the posterior) (eq. \ref{eq:free_energy_distribution}).
% F(S_t) = H(P(S_t|V_{t-1})) + D_{KL}(P(S_t|V_{t-1})||P(S_t|C_t))
\begin{equation}
\begin{split}
F(\mathbb{X}) &= H(P(\mathbb{X})) + D_{KL}(P(\mathbb{X})||Q(\mathbb{X})) \\
&= -\sum_iP(\mathbb{X}_i) ~ln P(\mathbb{X}_i) + \sum_{i}P(\mathbb{X}_i) \cdot ln \frac{P(\mathbb{X}_i)}{Q(\mathbb{X}_i)}
\end{split}
\label{eq:free_energy_distribution}
\end{equation}
% \textbf{In our interpretation (look if Howy has similar thoughts)} 
In our interpretation the difference between perception and production lies in the question which signal drives the updates in calculating free energy, the bottom-up signal or the top-down signal? When it comes to calculating free energy in \textbf{perception} the posterior distribution (for calculating surprise and the cross-entropy) is the top-down signal $P_{td}$ and the prior, i.e., the driving signal is the bottom-up signal $P_{bu}$, so that $F = H(P_{td}) + D_{KL}(P_{td}||P_{bu}))$.
In \textbf{production} the top-down signal $P_{td}$ becomes the driving signal for the free energy update $F = H(P_{bu}) + D_{KL}(P_{bu}||P_{td}))$. The deviation is then calculated with respect to $P_{bu}$ (see the yellow box in Figure~\ref{fig:model_update_summary}).
To update level beliefs, both posteriors will be combined to form the current level posterior $P_t$, in which the bottom-up signal will be combined similarly as in the identified microcircuitry for predictive coding \citep{Bastos:2012bd}. There, connections between cortical columns are mostly inhibitory. In this setting, the bottom-up and top-down posteriors in eq. \ref{eq:free_energy_distribution} play the role of predictions. This means that we can treat the differences in the bottom-up and top-down predictions as a prediction error and enter them into a Kalman filter. In this formulation, the Kalman gain is used to differentially weight bottom-up and top-down predictions and plays the role of a precision. Crucially, this precision is a function of the free energy computed at each level in the hierarchical model – such that a very high free energy (i.e. prediction error) emphasizes top-down predictions. 
We model this effect during \textbf{perception} as a top-down inhibitory influence on the bottom-up signal using a Kalman filter $P_t = P_{bu} + K_t (P_{td} - P_{bu})$ with a Kalman gain $K_t = \tfrac{F}{F + \pi}$ calculated from the level free energy $F$ and precision $\pi$ to integrate this filter into the model context (see below in eq. \ref{eq:soa_filter} for a more detailed description).
Again, the current driving signal can invert the belief update so that during \textbf{production} $P_t = P_{td} + K_t (P_{bu} - P_{td})$. The level posterior $P_t$ will be send to the next higher level in the hierarchy and to the next lower level. In the following time step the level posterior will be used as an empirical prior for calculating $P_{bu}$ and $P_{td}$ respectively.
Figure~\ref{fig:model_update_summary} examplifies our belief update scheme in Sequence level \emph{S}. There posteriors and relevant likelihoods are combined to calculate updated top-down and bottom-up posteriors necessary to calculate the free energy within level \emph{S} and the final belief update step which combines top-down and bottom-up posteriors.
For more details on our generative model hierarchy please refer to the Supplementary Materials.

% intention to act as additional signal is necessary
Following the assumption in active inference that overt action is basically action-oriented predictive processing \citep{Friston:2010ez}, we allow strongly predicted next actions of representations in the \emph{Sequence} level to be acted out. Without any constraints this leads to overt automatic \emph{enactment}, similar to the automatic imitation seen in patients suffering from echopraxia \citep{Ganos:2012fv}. To control this motor execution, we introduced a gating signal into the top-level of the hierarchy, which acts as a motor intention for a specific schema, including a strong boost of this schema's probability (acting as a trigger to act-out an abstract motor belief). This intention percolates down the hierarchy to boost associated representations and inform the intention to act. Once it reaches the lowest level of the hierarchy, the act to produce the motor representation will be allowed.
% ------------------------------------------------------------------------------------
\subsection{Motor coordination loop}
In active inference and the actual production of an action, predictions guide the minimization of free energy, the Motor level receives position ($x_i$) and timing ($\Delta_t$) goals from Sequence level $S$. The realization of the movement goal within the Motor Control level is left to an action-proprioception loop modeled as a dampened spring system. This models the angular movements of a single joint, which in our simulations will represent a writing pencil. A similar approach was used by \cite{Friston:2011cua} but instead of a system with two joints following one attractor-state at a time in attractor space we model just one joint which moves towards the spring system's point of equilibrium at the relative polar coordinate of the action goal.
To allow for smooth and curving trajectories that are similar to handwriting in spatial and temporal properties, we were inspired by work on dynamic movement primitives (DMP) that are used for modeling attractor behaviors of autonomous nonlinear dynamical systems with the help of statistical learning techniques \citep{Ijspeert:2013bt}. We will not make use of the DMPs ability to learn and reproduce trajectories, but will configure a dampened spring system similarly to a DMP and instead of applying a forcing term that activates the system's nonlinear dynamics over time we make use of an obstacle avoidance technique mentioned in \cite{Hoffmann:kk} which we adpoted and inverted its force to actually move towards the goal in a forcing function $g$ (see eq. \ref{eq:goal_forcing_function}). The reason for this is that when we simply applied the spring system to each goal sequentially, we would accelerate toward and slow down at the goal. To keep up the momentum we need to look ahead several goals $x_{i+3}$ (here 3 steps ahead) in the core spring system, but with a goal forcing function that sequentially tries to visit each goal $x_i$. $\alpha=25, \beta=6.25, \gamma=10$ and $\mu=\tfrac{1}{\pi}$ are constants that specify the behavior of the system. $\varphi$ is the angle to the goal (or its velocity, see eq. \ref{eq:dampened_spring_system}) and $y$ is the current position.
\begin{equation}
\begin{split}
\varphi &= \varphi_{x_i} - \varphi_{\dot{y}} \\
\dot{\varphi} &= \gamma ~\varphi ~e^{-\mu |\varphi|} \\
g &= (x_i - y) ~\dot{\varphi}
\label{eq:goal_forcing_function}
\end{split}
\end{equation}
\begin{equation}
\ddot{y} = \alpha (\beta (x_{i+3} - y) - \dot{y}) + g
\label{eq:dampened_spring_system}
\end{equation}
With this hierarchical model in place, we set out to investigate how a sensorimotor system can distinguish activations that stem from own actions from those arising due to the observation of an interaction partner's actions. In particular we are interested in how self-other distinction is realized within the system itself based on its prediction-based processing. This ability would be a prerequisite for the assumed dual use of the sensorimotor system (in perception and action) in dynamic social interaction. Note that, at a higher cognitive level, self-other distinction will have additional components based on sensory modalities like vision (e.g. seeing who the agent of an action is) or proprioception, as well as awareness of control of own actions (e.g. feeling or knowing that one is executing an action). Here we are interested in how a more fundamental sensorimotor basis for self-other distinction can contribute in correctly attributing a perceived sensory event to self-action or other-action -- a process that would be likely to underlie many of the these additional higher-level components. To that end, we extended HPBU to include a sensorimotor account of sense of agency (SoA), i.e., the sense that an action is self-generated.
%--------------------------------------------------------------------------------
\section{Self-other distinction and sense of agency}
How does the human brain distinguish between information about ourselves and others? Or to be more specific, how can we distinguish ourselves from others so that we do not falsely attribute an action outcome to ourselves? These questions are related to the general mechanisms that give rise to a sense of ``feeling of control'', agency, and ``self''. 

Reviews on the neural mechanisms of the SoA and the social brain \citep{David:2008dp,VanOverwalle:2009eh} show a strong overlap of differential activity during SoA judgement tasks with functional brain areas of the human mirror-neuron system and the mentalizing network. Especially noteworthy is TPJ (temporo-parietal junction) as a candidate to infer the agency of a social action, spanning areas STS (superior temporal sulcus) which mainly responds to biological motion, to IPL (inferior parietal lobule) which may respond to the intentions behind someone's actions, connecting to mPFC (medial prefrontal cortex) which probably hold trait inferences or maintains different representations of self- or other-related intentions or beliefs (please see the mentioned reviews for a more thorough analysis). 
Generally, a person's SoA is believed to be influenced through predictive and postdictive (inferential) processes, which when disturbed can lead to misattributions of actions as in disorders such as schizophrenia \citep{vanderWeiden:2015gk}. In Schizophrenia as a deficit of sensory attenuation, but also hallucinations, we can point to disfunctional precision encodings as a core pathology, i.e., the \emph{confidence} of beliefs about the world \citep{Adams:2013fi}. Precision as such is believed to be encoded in dopaminergic neuromodulation and can as such be linked to the sensory attenuation effects during the attribution of agency in healthy subjects \citep{Brown:2013eo}. We aim to identify mechanisms in order to model these processes and their integration into a combined SoA. 
% ------------------------------------------------------------------------------------
\subsection{Predictive process in sense of agency}
The \textit{predictive} process makes use of people's ability to anticipate the sensory consequences of their own actions. It allows to attenuate, i.e., decrease the intensity of incoming signals which enables people to distinguish between self-caused actions and their outcomes and those actions and outcomes caused by others. One account to model these processes is based on inverse and forward models to account for disorders of awareness in the motor system and delusion of control \citep{Frith:2000ksa}. This view suggests that patients suffering from such disorders of awareness can no longer link their intentions to their actions, that is they can become aware of the sensory consequences of an action, but may have problems to integrate them to the intention underlying the action, making it hard to ascribe actions to oneself or another agent. Research on schizophrenia has shown that reliable and early self-other integration and distinction is important not only for the correct attribution of SoA, but also for the correct attribution of intentions and emotions to others in social interaction \citep{vanderWeiden:2015gk}. \cite{Weiss:2011jn} also showed that there is a social aspect to predictive processes that influence SoA by comparing perceived loudness of auditory action effects in an interactive action context. They found that attenuation occured also in the interactive context, comparable to the attentuation of self-generated sound in an individual context.

Another aspect of the processing of differences between predictions and feedback from reality is the intrinsic robustness and invariance to unimportant aspects in the sensory input. The kinds of predictive processing hierarchies we talk about, that concern themselves with allowing ourselves to act in (and perceive) the ever-varying nature of our environments, are able to ignore or \emph{explain away} the prediction-errors that aren't surprising enough to lead to any form of adaptation. That this is also likely true for temporal prediction-errors was found in \cite{Sherwell:2016kz}, who using EEG saw significant N1 component suppression in predicted stimulus onset timings. In a predictive processing perspective this is possible through higher levels in the hierarchy correctly predicting the next lower level's state, and by taking into account its generative model's precision, minimize its prediction-error. Or as Clark states: ,,[...] variable precision-weighting of sensory prediction error enables the system to attend to current sensory input to a greater or lesser degree, flexibly balancing reliance upon (or confidence in) the input with reliance upon (or confidence in) its own higher level predictions.'' \citep[pp. 216]{Clark:da}. Consistent with this perspective is work by \cite{Rohde:2016jw}, who investigated if and in which cases we can compensate for sensorimotor delay, i.e., the time between an action and its sensory consequence. They find that if an error signal (a discrepancy between an expected and an actual sensory delay) occurs we recalibrate our expectations only if the error occurs systematically. This kind of temporal adaptation is a well studied finding (e.g., \cite{Haering:2015is} for sense of agency or \cite{Cunningham:2001gg} in motor control). What interests us for the here presented model of self-other distinction are the unexpected, unsystematic and sudden deviations that cannot be explained away easily. These are the temporal aspects of sensorimotor processing that we will focus on next. 
% ------------------------------------------------------------------------------------
\subsection{Postdictive process in sense of agency}
The \textit{postdictive} process relies more on inferences drawn after the movement in order to check whether the observed events are contingent and consistent with specific intentions \citep{Wegner:1999wy}, influenced by higher-level causal beliefs. One important aspect of this inferential process relies on the temporal aspects of action-outcome integration. It was shown that increasing action outcome delay decreases feeling of control \citep{Sidarus:2013be}. \cite{Colonius:2004iz} describe a model that can explain the improved response time in saccadic movements towards a target that is visually and auditorily aligned. Their \emph{time-window-of-integration} model serves as a framework for the rules of multisensory integration (visual, auditory, and somatosensory), which occurs only if all multimodal neural excitations terminate within a given time interval. In \cite{vanderWeiden:2015gk} this time interval of integration is taken as a solution to a problem posed in the classic comparator model of motor prediction. The brain needs to integrate action production signals with their predicted effects which can be perceived via multiple sensory channels (e.g., visual, auditory, proprioceptive, ...). This integration needs to account for the different time scales in which effects (or outcomes) of actions may occur. This is where the \emph{time-window-of-integration model} can help us explain the effect such integration can have. 

A point not taken into account by Colonius and Diederich was, how such an integrating mechanism knows how long it has to wait for all action outcomes to occur. \cite{HillockDunn:2012cm} investigated how these temporal windows for integration, which have been learned in childhood develop through life. They analyzed responses to a judgement task of a visual and an auditory stimulus to occur simultaneously in participants with ages ranging from 6 to 23 years. Their analysis showed an age dependent decrease in temporal integration window sizes. We hypothesize that a wider window of integration can be associated with unpredictability and greater variance in action outcome timings and that this integration window size decrease may be due to an adult person's experience advantage about effects their actions may have on their environment, or the mere better predictability of their full grown bodies. 

Such an integration of an intended action with its predicted consequences learned through associations between action events can lead to an interesting phenomenon, often reported in the SoA literature. In this phenomenon an integration can lead to the effect that predicted action consequences can be perceived to occur at the same time. This phenomenon is called \emph{temporal binding}, or also \emph{intentional binding} when the effects of an intended action are predicted and are perceived as occurring closer together as unintended actions that were merely observed in an unrelated party \citep{Haggard:2002fl,Haggard:2003df}. We have trained our sensorimotor hierarchy to allow the model of sense of agency to be able to predict also the \emph{timing} of its intended action outcomes. This way it is able to integrate action and consequence, while simultaneously evaluating the success of its outcome prediction and allowing the detection of unpredicted delays, which may be a cue for another agent's action and their action outcome or an outcome delayed by an unknown reason.
Being able to make such a distinction allows people to monitor, infer and distinguish between causal relations for own and other's behavior. 

% sum up
In sum, by and large there are two processes that can inform SoA and hence can help to distinguish actions of self and other in social interaction. A predictive process is based on (assumed or given) causes of the action, e.g., the motor command and utilizes forward models to predict the to-be-observed sensory events. A postdictive process works with features of an observed action outcome and applies higher-level causal beliefs and inferential mechanisms, e.g., a given intention to act or temporal binding, to test the consistency of the action outcome and infer a likely explanation.

% cue integration
How are these two processes integrated to inform SoA and what if their cues are unreliable? When disorders of SoA were first studied, the \textit{comparator} model was the first proposal concerning its underlying mechanism. This was soon questioned as the comparator model failed to account for external agency attributions. It was argued that its evidence has to be weighted and integrated with more high-level sources of evidence for sense of agency \citep{Synofzik:2008fl}. Such a weighting and integration of evidence cues was first studied by \cite{Moore:2009gd}, who found that external cues like prior beliefs become more influential if predictive cues are absent. Neurological evidence for a differential processing of cues that inform SoA comes from \cite{Nahab:2010js}, who found in an imaging study that there is a \textit{leading} and a \textit{lagging} network that both influence SoA prior to and after an action. The leading network would check whether a predicted action outcome is perceived, while the lagging network would process these cues further to form a SoA that is consciously experienced. Further, an EEG study found evidence for separate processing areas in the brain \citep{Dumas:2012fn}. There, predictive and postdictive cues were triggered in two tasks. One induced an external attribution of agency, while the other used a spontaneuous attribution condition. It seems that in order to generate SoA, both systems do not necessarily have to work perfectly together. Instead, there is evidence that the SoA is based on a weighted integration of predictive and postdictive cues based on their precision \citep{Moore:2012fw,Synofzik:2013hj,Wolpe:2014kh}. Furthermore, the fluency of action-selection processes may also influence self-other distinction because the success of repeatedly predicting the next actions seem to accumulate over time to inform SoA \citep{Chambon:2012hu,Chambon:2014cfa}. This action selection fluency aspect seems to contribute prospectively to a sense of agency, similar to a priming effect.
%--------------------------------------------------------------------------------
\subsection{Estimating sense of agency during action and perception} \label{sod}
% or "model of attribution of agency"?
During online social interaction, the sensorimotor system potentially gets involved in simultaneous action perception and production processes. Our goal is to investigate how our prediction-based HPBU model can cope with the social differentiation problem during such dual-use situations. To that end we integrate three cues that arise in predictive and postdictive processes, into a SoA for produced actions which will depend on the likelihood calculated in the Sequence level: In the predictive process, we calculate the likelihood of the perceived action sequence $s^\prime$, given the predicted action sequence $s_i \in S$ in the sequence comparison function $\iota(s^\prime,s_i)$. In the postdictive process, we have the intention to act and the delay in the action-outcome for temporal binding. This temporal binding depends on the predicted and perceived temporal delay of the predicted action, and the sequence level's precision. Precision in this context will stretch or sharpen the likelihood of temporal binding (see eq. \ref{eq:sequence_likelihood} and refer to Supplementary Materials for more details on $S$ and the sequence comparison function).
\begin{equation}
P(s^\prime|s_i) = \iota(s^\prime,s_i) \cdot e^{- \frac{(s^\prime_{\Delta t} - s_{i,\Delta t})^2}{2 ~\pi_S^2}}
\label{eq:sequence_likelihood}
\end{equation}
Following the evidence for a fluency effect that accumulates the repeated success in correctly predicting and selecting actions \citep{Chambon:2014cfa}, we model this accumulation of evidence. That is, we feed the likelihood of the current action of the intended sequence $s_I$ into a Kalman filter to estimate the agency (see eq. \ref{eq:soa_filter}). The Kalman filter estimates the agency $\hat{a}_t$ from the likelihood $P(s^{\prime}|s_I)$ and the previous agency estimate $\hat{a}_{t-1}$, the Kalman gain $K_t$ is calculated from the sequence level's free energy $F_S$ and precision $\pi_S$. This form of Bayesian belief updating is implemented as follows (noting that the conventional update of Kalman gain is replaced by a function of variational free energy):
\begin{equation}
\begin{split}
K_t &= \frac{F_S}{F_S + \pi_S} \\
\hat{a}_t &= \hat{a}_{t-1} + K_t (P(s^{\prime}|s_I) - \hat{a}_{t-1})
\label{eq:soa_filter}
\end{split}
\end{equation}
By allowing the agency estimate only to accumulate through this filter, strong fluctuations are dampened. Further, with the gain governed by precision and free energy the influence of the estimate will strongly depend on the success of previous predictions. 
The essential elements for the sense of agency of the perception-action loop are an intent for a specific action production, the correct prediction of the action, and it's timing, which was learned. In our generative model in the Sequence level, the prediction and evaluation of an action and it's timing are embedded in each sequence. The comparison function returns a value between 0 and 1. The intent for a specific action production essentially can be described as a high precision prediction that is very strong and stable over time. If it is then the case that such a high precision prediction is the driving signal and the probability for the predicted sequence stays low, the model free energy will be high. Our interpretation of this process and it's outcome is that either something unpredicted is influencing the action production or it is not the system's production at all that is perceived.
% ------------------------------------------------------------------------------------
\section{Simulations and Results}
To test the extended model's ability to solve the problem of the dual use of the sensorimotor system and of differentiating between self and other, we have simulated a number of scenarios and trained HPBU on a corpus of handwritten digits from 0-9. The handwriting corpus was previously recorded using a self-implemented app on an 6th generation iPad using the Apple Pencil as an input device. Each digit was recorded ten times by the same person. The learned temporal and spatial dynamics were learned by the system to allow for both the spatial and temporal information to influence the correct production and perception of a writing sequence. We leave the details of how HPBU learns new representations to a future publication.
\begin{figure}
\begin{center}
\includegraphics[width=0.84\textwidth]{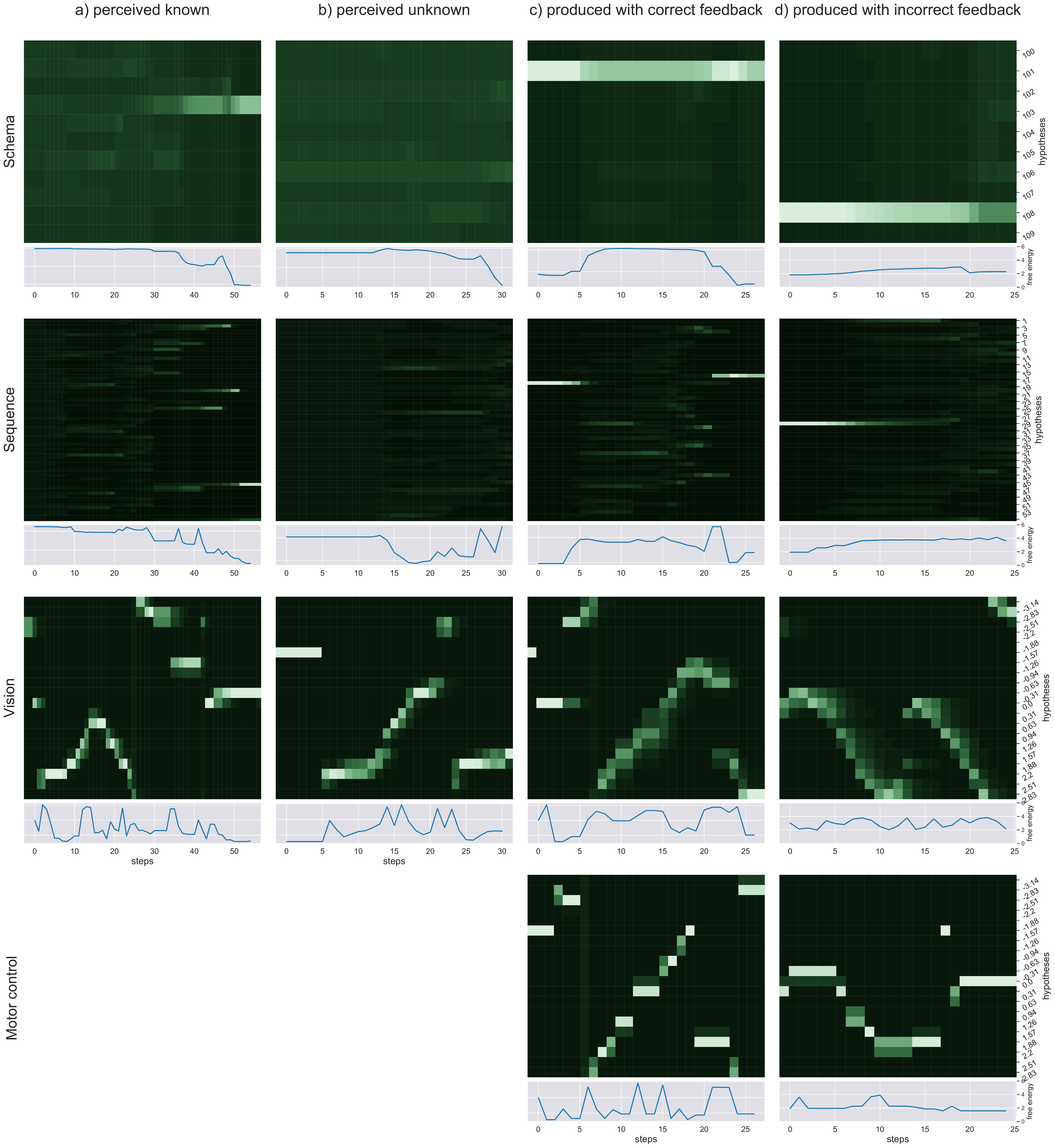}
\end{center}
\textbf{\refstepcounter{figure}\label{fig:scenario_probability_dynamics} Figure \arabic{figure}.}{We show the probability dynamics in the model as heatmaps over the probability distributions over time in the different levels of the HPBU (color coded from dark green to white from \emph{P=0} to \emph{P=0.6} for best differentiability). The different scenarios are \textbf{a) perceived known}: the perception of a known digit (here a 5), \textbf{b) perceived unknown}: the perception of an unknown digit (here a 4), \textbf{c) produced correct feedback}: the production of a digit by means of active inference (here the digit 9), and \textbf{d) produced incorrect feedback}: the production of a digit (here a 1), using the correct \emph{proprioceptive} feedback while the simulated production of another digit (here a 3) is received as \emph{visual} feedback. In addition the free energy dynamics for each level is shown. As one can see, the visual input clearly influences the perception of sequences and schemas of sequences in higher levels, thereby minimizing free energy over time. Also, during production the belief created in the schema representing the digit 9 percolates down the hierarchy, activating and acting out a selected sequence. In scenario d) the sequential activation is shown for producing one digit in the motor control level, while seeing the activation dynamics for visually perceiving another digit in the vision level. The resulting confusion is immediately visible in sequence level, and reduces in schema level by settling on a lower probability for the preferred hypothesis.}
\end{figure}
We simulated the production and perception of hand written digits in four scenarios and over time recorded the respective changes in free energy, probability distribution and the sense of agency estimate. First, we simulated the pure perception of a) a known sequence of actions to write a digit and b) an unknown way to write a digit. Then, we simulated the writing of a digit with c) correct visual and proprioceptive feedback. To simulate scenarios in which the distinction of self and other is relevant d) we had the model write a digit, gave the correct temporal and spatial \emph{proprioceptive} feedback, but gave temporal and spatial \emph{visual} feedback of a different (but known) way to write another digit. Figure~\ref{fig:scenario_probability_dynamics} shows the probability dynamics in the model as heatmaps of the probability distributions over time in the different levels of the HPBU hierarchy for the different scenarios. The heatmap color codes the probability distributions from dark green to white for best differentiability. Each heatmap shows the level's posterior probabilities after beliefs are updated bottom-up and top-down. Simultaneously, the free energy dynamics are plotted for each level, showing the level of adaptation and model evidence given the current state of the system. The bottom row represents activity in the Motor Control level, which is non-existent in the purely perceptual scenarios a) and b).

In scenario a) the heatmaps show nicely how the Vision level perceives the different movement angles over time. Simultaneously, evidence for the sequence hypotheses accumulates slowly with each new salient visual feature, finally leading to at first only a limited number of probable sequence representations and finally to a single one. Also, the schema hypotheses accumulate evidence even more slowly, predicting the underlying sequences. Schema level predictions have a strong influence on the Sequence level which is most evident in the final Sequence level distributions in which only a number of sequences are still probable and most of them belong the the most probable schema hypothesis. In scenario b) no such evidence accumulation is present, as the shown digit was not known.

Between scenarios c) and d) the most important difference can be seen between Motor Control and Vision levels of each. There, the posterior probabilities shown in the Motor Control level heatmap should be similar to the Vision level heatmap, but is only so in scenario c), where production and perception align. To that effect, the heatmaps of Sequence and Schema levels show how evidence for predictions cannot be met in scenario d), but are mostly met in scenario c). Interestinly, in scenario c) evidence for the first predicted sequence is not met at some point, so that another viable sequence hypothesis from the same schema hypothesis becomes active after some time. This pertubation may be due to spring dynamics in the Motor Control.
\begin{figure}[t]
\begin{center}
\includegraphics[width=0.6\textwidth]{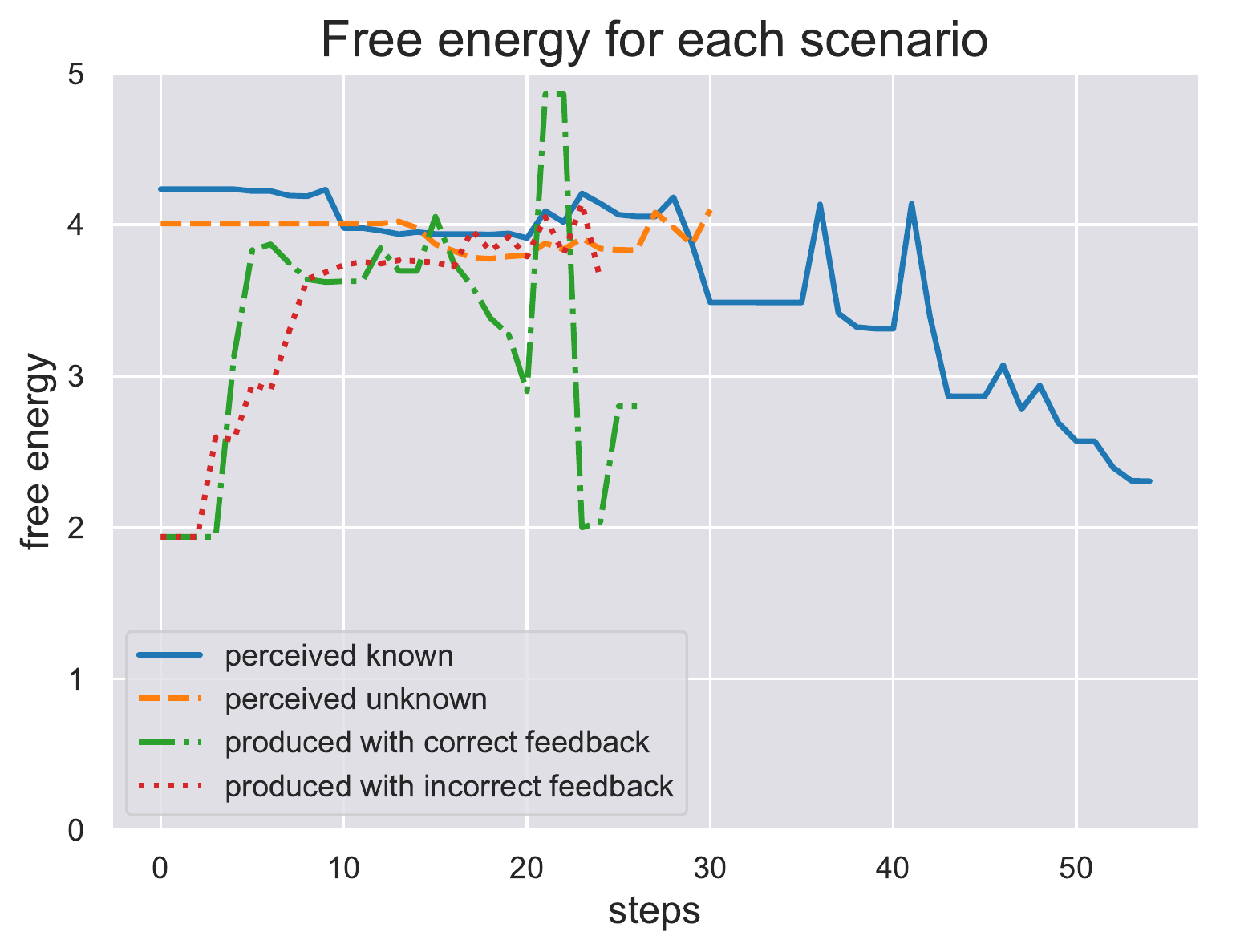}
\end{center}
\textbf{\refstepcounter{figure}\label{fig:scenario_free_energy} Figure \arabic{figure}.}{Here we plot the free energy of the sequence level during each test scenario. The different sequences perceived or produced in each scenario have different lenghts. First, the perception of a known and unknown writing sequence clearly shows a difference in that free energy is minimized during perception of the former sequence, but not the later. Second, the production of a writing sequence also minimizes free energy as it would in active inference as long as the sequential production is successfull, i.e., the temporal and spatial prediction of sequential acts are met. In production scenario d) with incorrect feedback these predictions are met only proprioceptively but not visually. This shows in the plot where free energy cannot be minimized through active inference when its action driving predictions are met with contradicting visual feedback.}
\end{figure}
Having a look at the free energy throughout the hierarchy can give us an idea how well the model is able to find explanations for the perceived input. Strong fluctuations can be a clue to highly irregular input, e.g., in the vision level where bottom-up sensory evidence and top-down predictions can change rapidly. To get a better idea for the model's explanatory power in our case, i.e., the perception and production of sequences of writing digits, the sequence level's free energy dynamics can quickly respond to unpredicted input but still receives predictions from the schema level to inhibit the most unlikely explanations.

In Figure~\ref{fig:scenario_free_energy} we plot the free energy dynamics of the sequence level during each test scenario. As the different writing sequences are of different length, so are the model responses shown in this figure. The free energy plots show a successful minimization in scenario a) where the model perceives a known stimulus. It appears to quickly choose the correct hypotheses right from the beginning. In contrast, the model cannot successfully minimize its free energy in response to perceiving the unknown stimulus in scenario b). Interestingly, in response to the production in scenario c) the model first minimizes free energy, acting out a chosen sequence, but free energy spikes during production as some dynamics during motor control have not been predicted. In the end the sequence hypothesis switches to a similar sequence from the same schema and free energy minimization continues. In scenario d) the production and proprioceptive feedback of one written out digit is met be the visual temporal and spatial feedback of another digit. The free energy can thus not be minimized as predictions from schema and sequence levels are not met by correct visual feedback.
\begin{figure}[t]
\begin{center}
\includegraphics[width=0.6\textwidth]{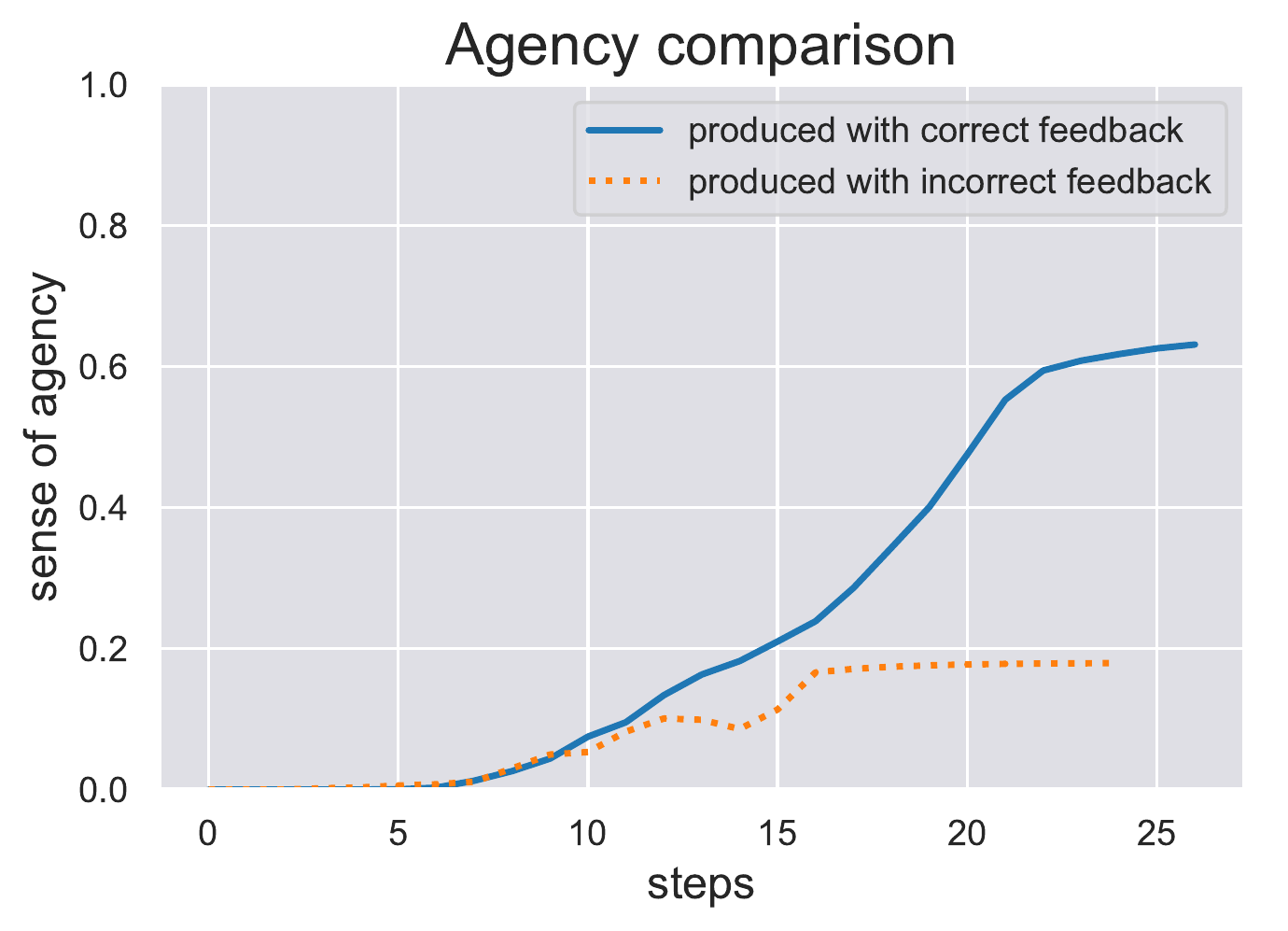}
\end{center}
\textbf{\refstepcounter{figure}\label{fig:scenario_soa} Figure \arabic{figure}.}{Here we see the SoA estimate dynamics produced by the HPBU in the production scenarios c) and d) with correct and incorrect feedback, respectively. Production and perception in the later scenario didn't take as long as in the scenario c). In both scenarios the SoA estimate rises up to a certain point but it remains at a low level of \emph{0.2} in scenario d) where predictions of produced actions are met with contradicting visual feedback.}
\end{figure}
We are interested in the SoA estimate calculated during active inference in order to see how well the model can be sure if it actually has acted on the world as expected. We ran two test scenarios, c) and d), where the model produced a writing sequence so only those are interesting to inspect its inferred SoA estimate. In Figure~\ref{fig:scenario_soa} we plot the SoA estimate dynamics produced in the scenarios c) (production with correct feedback) and d) (production with incorrect feedback). As one can see, the later scenario didn't take as long as the other to complete and in both scenarios the SoA can accumulate up to a certain point, but remains at a rather low level of \emph{0.2} in the end of scenario d), because temporal and spatial predictions about produced actions are met with contradicting visual feedback. In contrast, the SoA estimate can quickly accumulate quite high in scenario c), despite the spring dynamics during motor control, that lead to a spike in free energy and a switch to a different sequence hypothesis.
% ------------------------------------------------------------------------------------
\section{Discussion}
In dynamic social scenarios of concurrent perception and production of actions, where sensory events can originate either from own or from other's actions, a "dual-use" sensorimotor system that is presumed to be involved in both processes has a challenging task. Specifically, it needs to distinguish in its processing between self-action and other-action, subserving its functions for action execution and recognition. We have presented a model called empirical belief correcting hierarchy based on active inference and extended it with mechanisms of a SoA that enables the judgement -- already at these levels of sensorimotor processing -- that an action is self-generated. In line with current views on SoA, this extension consists in a dynamic integration of predictive and postdictive cues which is embedded in the likelihood function that matches temporal and spatial aspects of perceived action sequences with those of known sequences. 

The resulting dynamics of our simulation scenarios show the successful minimization of free energy during perception and production (see scenarios a) and c) in Figure~\ref{fig:scenario_probability_dynamics}). There, over time the different possible schemas and their sequences are either considered as possible explanations for the perceived stimuli, or the sequences and actions are chosen for the intended schema to be acted out. In both cases free energy is minimized. In the perception case the minimization is due to the selection of possible explanations, which then can correctly predict further aspects of the action. In the production case the correct selection and prediction of action effects that allow for the intended schema to be a possible explanation leads to the free energy minimization. 
In both cases free energy is not merely an epiphenomenon of the system, i.e., an acausal property that results from the system dynamics. Free energy is part of the belief update scheme that integrates top-down and bottom-up posteriors, by influencing the Kalman gain that allows for the signal to influence the prior. Dependend on the current mode, whether free energy is minimized through action or perception, the top-down and bottom-up posteriors take different roles (prior or signal) in this update scheme. In addition, free energy also similarly to the belief update scheme, controls the precision or Kalman gain controlling the accumulation of the continuous SoA estimate. Together, this qualifies as active inference. 
Furthermore, the implementation we present here can be described as an example of a ,,deep temporal model'' \citep{Friston:2017fj}. For example, during bottom-up inference the model accumulates evidence over time to consider different possible explanations, and top-down a state at any level of the model hierarchy can entail a sequence of state transitions in it's next lower level.

A possible line of criticism is that we control for our simulation scenarios which \emph{mode} the model finds itself in, through our intention signal. The intention signal is received by the Schema level and tags one of its hypotheses for production in active inference. Then, using additional connections to other levels in the hierarchy the intention is spread, giving and maintaining a boost of probability to associated hypotheses in those levels. In our implementation we found this maintained probability boost to be necessary for the inference to inhibit sudden switches to other explanations (and action sequences). This maintained probability or activation is similar to activity that is maintained during attentional tasks in area MT \citep{Treue:1999ep}, i.e., where monkeys were tasked to follow a single moving visual stimulus while other movements were also on display. Naturally, the probability of intended sequences should be maintained through the correct prediction of actions. In reality small differences in the feedback to early actions of an action sequence can lead to increased probabilities of more similar sequences that not always belong to the intended schema. Even with the intention signal in our production scenario c), small unpredicted pertubations in the spring dynamics of Motor Control led to a switch to a similar sequence from the intended schema. 
We argue that while such an intention signal seems artificial in our implementation, it is necessary in a limited hierarchy, such as ours. It may well be the case that in an extended hierarchy the maintained boost in probability may be provided by higher levels through appropriate priors.

During scenario b) an unknown writing sequence was perceived and free energy could not be minimized, although the Schema level seems to settle for one possible hypothesis. In the Sequence level no viable hypothesis can be found and free energy stays high. In our implementation this maintained inability to minimize free energy is a reliable signal for the need to extend upon possible hypotheses to choose from. As such, Bayesian models lack the ability to extended upon their hypothesis space for the reason that renormalization becomes necessary. In a hierarchical model such an additional sequence would have to be associated with a corresponding schema in the Schema level in order to embed and associate it with similar sequences, if there are any. For a sequence such as which was used in scenario b) no similar sequence could be found simply because the model was not trained on these. Specifically, from all digits seens during learning, all writing sequences of \emph{fours} have been excluded. This way, no similar sequence could be found which could have explained the perceived stimuli. For HPBU learning this not only means that another sequence should be learned, but also the addition of a new schema becomes necessary. The learning strategy is still in its early stages, thus we do not go into much detail here and leave this for future work. 
% Although, we have included some remarks on it in our supplementary material. (\emph{REFERENCE SUPPLEMENTARY MATERIAL})

The results of our simulations show (see Figure~\ref{fig:scenario_soa}) that the cue integration mechanism, which is integrated into the likelihood function of the Sequence level, supports results reported in the literature. First, it is sound with regard to results where the reliability (here: precision) of the predictive process was reduced and the system put more weight on postdictive processes, conforming evidence for a weighted integration based on the cues' precision \citep{Moore:2012fw,Synofzik:2013hj,Wolpe:2014kh}. This aspect can be observed in scenario d), where the SoA estimate increases slowly even though a completely different digit is being perceived visually. This may be due to the fact that when drawn simultaneously, a perceived 3 and a drawn 1 start similarly, despite the roundness in the trajectory of a three. When either timing or spatial predictions are met to a degree they can accumulate. In our simulation it is then only the lower, second curved trajectory of the 3 that is in total contrast to the trajectory of the 1, which finally prohibits further accumulation of agency.
Second, our results are in line with a fluent correct prediction of actions \citep{Chambon:2014cfa}. The accumulation of the SoA estimate over time is done using a Kalman Filter, which depending on the current free energy and precision of the sequence level, filters out strong fluctuations. The more accurate the hierarchy's predictions the faster the uptake of SoA evidence (positive and negative). Looking at scenario c), this may be the reason why the later spike in free energy can stop the further increase of SoA, which plateaus in the end. Finally, even though the cue integration model is flexible with regard to the precision of predictive and postdictive cues, scenario d) showed that a false attribution of SoA is not likely when both cues show no signs of agency. \\
Other than \cite{Friston:2011cua}, who rely heavily on proprioceptive information we allowed for visual information to solely drive motor coordination. We closed the motor coordination loop using a direct connection that is used by the Motor Control level to inform the Vision level when it is done coordinating actions to reach a subgoal. Vision level will then check if visual information can confirm the movement and close the motor coordination loop by sending the information to Sequence level. Using a motor coordination loop that heavily depends on visual information allowed us to easily trick the model into doubting its own action production, by feeding it visual information of a different writing sequence, as in scenario d), the production with incorrect feedback. \\
Overall, the results reported here show that the model's attribution of SoA to its own action outcomes is affected in relatively realistic and, more importantly, differentiated ways when receiving different simultaneous perceptual information. This suggests, at least to some extent, that the motor system can play an important role in realtime social cognition as proposed by \cite{SchutzBosbach:2006bi}. Still, the literature on the social brain suggests that motor cognition as well as the distinction of self and other are influenced by higher-level processes, causal beliefs, and the mentalizing network. We agree that the interplay between mentalizing and mirroring needs to be incorporated to meet the demands of truly social systems in interaction scenarios with multiple agents. Also, we need to mention that since this is a deterministic model with a representation size that can be handled without any need for sampling we report purely qualitative descriptions of our simulation results. Also, we have disabled learning during our simulation runs, so that every simulation will return the same results, without any variance. 
A line of criticism might be that from the mere attribution of SoA to an action in our HPBU model we can hardly deduce more than a kind of tagging of an action within the hierarchy. But we think that when the current conceptual model is embedded within an extended HPBU model that will also cover the functionality of mentalizing areas, an agency attribution will help to confirm motor beliefs attributed to a prospective model of the self or distinguish its own actions from those of an interaction partner.  

\section{Conclusion}
We have presented Hierarchical Predictive Belief Update (HPBU) which models a predictive sensorimotor hierarchy. We integrated HPBU with a model of the sense of agency, which allows to dynamically integrate cues for sense of agency (SoA). This SoA attribution to an action enables the judgement that an action is self-generated. At the core of this modeling approach is the assumption that we strongly rely on the predictability of our own body to be able to differentiate between our own and other's actions through the information gathered especially from the unpredictability of others. The functional simulation helped to evaluate that the identified mechanisms in the model are sufficient to correctly infer own actions from feedback to its action production.

Furthermore, we presented simulation results of different scenarios of perception and production. We discussed how the model's dynamics and how it is able to minimize its free energy in each situation. In two scenarios we simulated action productions in which the SoA could be inferred and compared them to the literature on SoA and the influence of the motor system on social cognition. This comparison suggests that HPBU can correctly attribute SoA for its own actions, using a flexible integration of predictive and postdictive cues while integrating the evidence over time.

We made the model's sequence coordination independent of direct proprioceptive feedback by closing the motor coordination loop via the Vision level, without direct Motor Control level to Sequence level connection. This may have broad implications for the coordination and association of actions in the world with distal effects that do not directly or necessarily feed back into motor coordination. One example is the association of a switch on the wall with the distal effect of switching on the light on the ceiling. A more social example is the association of an action on another agent like smiling, with the effect of influencing that agent's emotional state. We want to explore this exciting possibility in future work.

The presented work is part of a research project investigating computational mechanisms underlying the intra-personal interplay between mentalizing and mirroring and the inter-personal coordination between interaction partners. We believe that computational cognitive modeling such as ours can be informative to the investigation of social cognitive processes which neuroscience is currently not yet able to elucidate, but where an analysis of the behavior of cognitive models based on findings from the neuroscientific literature can help to shed some light.

In future work, we want to improve our setup by making use of the information provided by the present model of self-other distinction to inform computational models of higher-level cognition through an interplay with the mentalizing system, in the process helping it to grasp another agent's intentions and beliefs.
In a first step toward this belief attribution we explored a rule based mentalizing model in previous work \citep{Kahl:2015uy}. Here, we lay the foundation to integrate a mentalizing model (higher-level social cognition) with the HPBU model on the common basis of active inference. This way, the mentalizing model will naturally be informed by a model of the sensorimotor system, while influencing it through its predictions. We conjecture that this interplay between the mentalizing and sensorimotor systems can yield the distinction between one's own and an interaction partner's beliefs needed in social interaction, where informed reciprocity is the key to efficient and successful communication.

\section*{Conflict of Interest Statement}
The authors declare that the research was conducted in the absence of any commercial or financial relationships that could be construed as a potential conflict of interest.

\section*{Author Contributions}
S.Kahl designed the model and the computational framework and performed the computational simulations. Both S.Kahl and S.Kopp contributed to the final version of the manuscript. S.Kopp supervised the project.

\section*{Acknowledgments}
This research/work was supported by the Cluster of Excellence Cognitive Interaction Technology 'CITEC' (EXC 277) at Bielefeld University, which is funded by the German Research Foundation (DFG).

\vskip 0.2in
\bibliography{bibliography}
\bibliographystyle{apalike}

\pagebreak
% \widetext
\begin{center}
\textbf{\large Supplemental Materials: A predictive processing model of perception and action for self-other distinction}
\end{center}
%%%%%%%%%% Merge with supplemental materials %%%%%%%%%%
%%%%%%%%%% Prefix a "S" to all equations, figures, tables and reset the counter %%%%%%%%%%
\setcounter{equation}{0}
\setcounter{figure}{0}
\setcounter{table}{0}
\setcounter{section}{0}
\setcounter{page}{1}
\makeatletter

\section{Domain and dynamical system}

Our hierarchical generative model can be expressed in terms of functions that map the internal state space from one level $L_X$ in the hierarchy to the domain of it's next lower level $L_{X-1}$, i.e., $L_{X-1} = l(L_{X})$, with \emph{l} as a placeholder for a specific mapping. The lowest level in the hierarchy will then map to the environmental state space $\varphi$.\\

In the highest level \textbf{Schemas} map to it's next lower level \textbf{Sequences} with $s: C \mapsto S $ where C cluster Sequences.

Level \textbf{Sequences} maps to it's next lower level \textbf{Vision} with $v: S \mapsto V$ where S consists of sequences of states from V. Then similarly, Sequences maps to the Motor level $M$ using function $m: S \mapsto M$. 

The lowest levels of the hierarchy are described by M (\textbf{Motor}) and V (\textbf{Vision}). Here, only the Motor level can directly influence the environmental state space through $m: M \mapsto \vartheta$ where $M$ represents movements in $\vartheta$.

We define the environmental state space in terms of a dynamical system of $\Omega = (X, T, \varphi, \vartheta)$, with space $X = \mathbb{R}^2$ and discrete time $T = \mathbb{N}$. $\varphi: X \times T \rightarrow X$ is a function of discrete movements over time and $\vartheta: T \rightarrow X$ is a function of movements in the state space (updating positions).

As an extension, our hierarchical generative model can be described as \emph{active} in the sense that each level maps external ($L_e$) to internal states ($L_i $) to minimize entropy. What is described as external depends on where in the hierarchy the level is situated. In the lowest level, i.e., Motor and Vision levels, the external states describe the sensory states of the system. Entropy minimization can be described as a function of $h(L_e,L_i)$ which maps external states onto internal states in a way that minimizes entropy (see eq. \ref{eq:entropy}).
\begin{equation}
\begin{split}
h(L_e,L_i) &= \argmin_{l \in L_i} H(L_e|l) \\
H(L_e|L_i) &= -\sum_{l \in L_e} p(l|L_i) ~ln ~p(l|L_i).
\label{eq:entropy}
\end{split}
\end{equation}

\section {Hierarchical generative model}

In the following the specific levels of the generative model (C, S, M, V, x) are described (see joint distribution in eq. \ref{eq:joint_distribution}). These are sequentially updated, i.e., they are updated in sequence updated from it's next higher and next lower levels, learning to represent and produce the states in the next lower level and the environment with $x = \varphi(x,t)$.
\begin{equation}
\begin{split}
P(C, S, M, V, x) &= P(C) \cdot P(S|C) \cdot P(V|S) \cdot P(M|S) \cdot P(x|M)
\label{eq:joint_distribution}
\end{split}
\end{equation}

\subsection{Schema level}

The Schema level $L_C = (C, F_C, \pi_C)$ (see eq. \ref{eq:schema_gen_model_definitions}) contains a discrete probability distribution over discrete states $C = \{c_1, \dots , c_n\}$, the free energy over the probability distribution $F_C$ and the precision over the probability distribution $\pi_C$, calculated as the reciprocal of the variance $\sigma^2$ over $P(C)$.
Each schema $c_i$ clusters similar sequences $\mathbb{S} \subseteq S$ from Sequence level $L_S$, and the schema's prototype $c_i^{\eta}$ over it's sequences.
\begin{equation}
\begin{split}
C &= \{c_i, \dots , c_n\} \\
F_C &= F(C) \\
\pi_C &= \tfrac{1}{\sigma^2(P(C))} \\
c_i &= (\mathbb{S}, c_i^{\eta}) \\
D_{\iota}^{n \times n} &= 
	\begin{bmatrix}
	\iota(s_1,s_1)& \dots& \iota(s_n,s_1)\\
	\vdots& \vdots& \vdots\\
	\iota(s_n,s_1)& \dots& \iota(s_n,s_n)
	\end{bmatrix}
	 \Bigl\lvert ~\forall s_{i,j} \in \mathbb{S} \\
c_i^{\eta} &= s_j ~| ~j = \argmin_{k} \sum_i^k (D_{\iota})_{i,k}
\label{eq:schema_gen_model_definitions}
\end{split}
\end{equation}
The schema prototype $c_i^\eta$ is calculated over the distance matrix over all it's sequences $D_\iota^{n \times n}$ (for distance function $\iota$ see eq. \ref{eq:sequence_distance}), finding the sequence with the minimal summed distance to all other sequences. This effectively defines the \emph{median} sequence clustered into the schema as the schema's prototype (similar as in \emph{k-medoid clustering}).

The generative model (eq. \ref{eq:schema_gen_model}) calculates it's posterior using the soft evidence ,,all things considered'' method over sequences given the schema \emph{cluster} it belongs to (for more information, please see \cite[chapter 3.6.1]{darwiche2009modeling})).
\begin{equation}
P_{bu}(C) = P_t(C) = \sum_{s_i \in S} P_{t-1}(s_i) \cdot P_t(c_i|s_j) ~~| ~c_j \in C
\label{eq:schema_gen_model}
\end{equation}

\subsection{Sequence level}

The Sequence level $L_S = (S, F_S, \pi_S)$ (see eq. \ref{eq:sequence_gen_model_definitions}) contains a discrete probability distribution over discrete states $S = \{s_1, \dots , s_m\}$, the free energy over the probability distribution $F_S$ and the precision over the probability distribution $\pi_S$.
Each sequence $s_i$ contains a tuple of observed movements $(o_1, \dots , o_k)$ in polar cooridnates at time $t \in T$ with $o_i = (\theta, r)$, and the time delay between observations $(\Delta_2, \dots , \Delta_k$), with $\Delta_it = t_i - t_{i-1}$.
\begin{equation}
\begin{split}
F_S &= F(S) \\
\pi_S &= \frac{1}{\sigma^2(P(S))} \\
S &= \{s_1, \dots, s_m\} \\ 
s_i &= ((o_1, \dots , o_t), (\Delta_2, \dots , \Delta_t))
\label{eq:sequence_gen_model_definitions}
\end{split}
\end{equation}
The difference between sequences $\iota(s_i, s_j)$ is calculated using the so-called ,,alphabetic Jensen-Shannon Distance'' (aJSD, \cite{Mateos:2017bp}), which first discretizes timeseries data in a combined probability space, on which then the Jensen-Shannon Distance can be applied (see the algorithm description in eq. \ref{eq:sequence_distance}). First, both $s_i$ and $s_j$ are converted into binary sequences $b_i$ and $b_j$, which are then both split into consecutive words $w_t(b)$ of word length $d=3$ with step size $\tau=1$ and collected in alphabets of words $W_i$ and $W_j$. Then, probability distributions $P^{W_i}$ and $P^{W_j}$ are formed from word frequencies for each alphabet. The Jensen-Shannon Distance $D_{JS}$ is calculated over these word frequency probability distributions.
\begin{equation}
\begin{split}
\varrho_t &= 
    \begin{cases}
      1,& \text{if } o_{t+1}> o_t \\
      0,& \text{if } o_{t+1}\leq o_t
    \end{cases} ~\Bigl\lvert ~\forall o_{t+1},o_t \in s_l \\
b_l &= (\varrho_1, \dots, \varrho_t) \\
w_t^{(d,\tau)}(b_l) &= (\varrho_{t-(d-1)\tau}, \dots , \varrho_t) ~| ~t \geq (d-1)\tau, \varrho_t \in b_l \\
W^{(d,\tau)}(b_l) &= \{w_1^{(d,\tau)}, \dots , w_t^{(d,\tau)}\} \\
P^{W_l} &= P(W_l^{(d,\tau)}) \\
\iota(s_i,s_j) &= D_{JS}(P^{W_i}||P^{W_j})
\label{eq:sequence_distance}
\end{split}
\end{equation}
We have two posterior distributions in the generative model for $L_S$ resulting from a bottom-up and a top-down update process ($P_{bu}(S)$, and $P_{td}(S)$ respectively). The top-down update for $P_{td}(S)$ calculates the posterior from a mixture of experts in Schema level $L_C$ (see eq. \ref{eq:sequence_gen_model_td}). 
\begin{equation}
\begin{split}
P_{td}(S) &= \sum_{c \in C} P(S|c) \cdot P(c)
\label{eq:sequence_gen_model_td}
\end{split}
\end{equation}
For the bottom-up update we need to calculate the sequence probability given observations represented in Vision level $L_V$.
$L_V$ represents only singular observations of movement, so in order to calculate their likelihood we collect observations in a temporally growing sequence $s^\prime = (o^\prime_{1}, \dots , o^\prime_\tau)$. We calculate the likelihood for each sequence $P(s^\prime|s_i)$ in which the sequence difference $\iota$ is weighted by an exponential factor, which calculates the temporal precision of the observed state.
This is in effect comparable to calculating a joint probability for all observation events $P(o^\prime_{1}, \dots , o^\prime_\tau|s_i)$.
Calculating the posterior then simply is a Bayesian inversion (eq. \ref{eq:sequence_gen_model_bu}).
\begin{equation}
P(s^\prime|s_i) = \iota(s^\prime,s_i) \cdot e^{- \frac{(s^\prime_{\Delta t} - s_{i,\Delta t})^2}{2 ~\pi_S^2}}
\end{equation}
\begin{equation}
P_{bu}(S) \approx P(S|o_1, \dots, o_{\tau}) = \frac{P(o_1, \dots, o_{\tau}|S) ~P(S)}{P(o_1, \dots, o_{\tau})}
\label{eq:sequence_gen_model_bu}
\end{equation}
Since a sequence will be of no use as a prediction for Vision level $L_V$ we need to obtain the next observation probability from the predicted sequence (if there is any) given the predicted sequence and prior observations (see eq. \ref{eq:sequence_event_prediction}). The resulting distribution is compatible with representations of Motor level $L_M$ and is used accordingly as $P(M|S)$.
\begin{equation}
P(V|S) \approx P(o_{\tau+1}|o^\prime_1, ... , o^\prime_\tau, s_i) = \frac{P(o^\prime_1, ... , o^\prime_\tau, o_{\tau+1}|s_i)}{P(o^\prime_1, ... , o^\prime_\tau|s_i)} ~\Bigl\lvert ~\forall o_{\tau+1} \in V 
\label{eq:sequence_event_prediction}
\end{equation}

\subsection{Vision level}

The Vision level $L_V = (V, F_V, \pi_V)$ (see eq. \ref{eq:vision_gen_model_definitions}) contains a discrete probability distribution over discrete states $V = \{v_1, \dots , v_i\}$ representing perceived movement angles, the free energy over the probability distribution $F_V$ and the precision over the probability distribution $\pi_V$. Each $v_i$ represents a movement perceivable by the model.
\begin{equation}
\begin{split}
V &= \{v_1, \dots, v_i\} \\
% V &= \{\tfrac{-\pi (10-i)}{10} ~| ~i \in I\} \cup \{\tfrac{\pi i}{10} ~| ~i \in I\}, ~\text{with} ~I=[0, 9] \\
F_V &= F(V) \\
\pi_V &= \tfrac{1}{\sigma^2(P(V))} 
\label{eq:vision_gen_model_definitions}
\end{split}
\end{equation}
The level calculates a top-down updated posterior $P_{td}$ (see eq. \ref{eq:vision_gen_model_td}) and a bottom-up updated posterior $P_{bu}$ (see eq. \ref{eq:vision_gen_model_bu}). $P_{td}$ is updated from a mixture of experts in Sequence level $L_S$. 
\begin{equation}
\begin{split}
P_{td}(V) = \sum_{s \in S} P(V|s) \cdot P(s)
\label{eq:vision_gen_model_td}
\end{split}
\end{equation}
The bottom-up update is a mapping from input state space $\varphi(x,t)$ to the model's movement reportoire using a gaussian likelihood function for each available movement $v_i \in V$, given $\sigma=0.1$.
\begin{equation}
\begin{split}
P(\varphi(x,t)|v_i) &= e^{-\frac{(\varphi(x,t)-v_i)^2}{2\sigma^2}}\\
P_{bu}(V) &\approx P(V|\varphi(x,t)) \propto P(\varphi(x,t)|V) ~P(V)
\label{eq:vision_gen_model_bu}
\end{split}
\end{equation}

\subsubsection{Salience detector}
To detect a salient input event we calculate the model free energy given two consecutive input events from state space $\varphi(x,t)$ Vision level $V$. If $F_t(V) > F_{t-1}(V)$ the updated model was not able to correctly predict the current input event. This is a salient event which will be communicated to the next higher level $L_S$.

\subsubsection{Oculocentric coordinates in Vision}
Salient movements observed in the environment are mapped to an oculocentric coordinate system before they are sent to the next higher level $L_S$. This way we map salient events in the environment to internal representations that are reusable as low-level action goals for Vision and Motor levels.
The oculocentric coordinates are relative polar coordinates ($\phi$, r) of the visual field, which when seen in sequence are similar to saccadic eye movements. In Motor level $L_M$ these coordinates will guide action in the form of movement goals.

\subsection{Motor level}

The Motor level $L_M = (M, F_M, \pi_M)$ (see eq. \ref{eq:vision_gen_model_definitions}) contains a discrete probability distribution over discrete states $M = \{m_1, \dots , m_i\}$ representing movement angles, the free energy over the probability distribution $M_F$ and the precision over the probability distribution $\pi_M$. Each $m_i$ represents a movement perceivable by the model.
\begin{equation}
\begin{split}
% M &= \{\tfrac{-\pi (10-i)}{10} ~| ~i \in I\} \cup \{\tfrac{\pi i}{10} ~| ~i \in I\}, ~\text{with} ~I=[0, 9] \\
M &= \{m_1, \dots, m_i\} \\
F_M &= F(M) \\
\pi_M &= \tfrac{1}{\sigma^2(P(M))} \\
\label{eq:motor_gen_model_definitions}
\end{split}
\end{equation}
The level again calculates a top-down updated posterior $P_{td}$ (see eq. \ref{eq:motor_gen_model_td}) and a bottom-up updated posterior $P_{bu}$ (see eq. \ref{eq:motor_gen_model_bu}). $P_{td}$ is updated from a mixture of experts in Sequence level $L_S$. 
\begin{equation}
\begin{split}
P_{td}(M) = \sum_{s \in S} P(M|s) \cdot P(s)
\label{eq:motor_gen_model_td}
\end{split}
\end{equation}
The bottom-up update is a mapping from input state space $\varphi(x,t)$ to the model's movement reportoire using a gaussian likelihood function for each available movement, given $\sigma=0.1$.
\begin{equation}
\begin{split}
P(\varphi|m_i) &= e^{-\frac{(\varphi(x,t)-m_i)^2}{2\sigma^2}}\\
P_{bu}(M) &\approx P(M|\varphi(x,t)) \propto P(\varphi(x,t)|M) ~P(M)
\label{eq:motor_gen_model_bu}
\end{split}
\end{equation}

\subsubsection{Motor coordination in active inference}
To allow for smooth and curving trajectories that are similar to handwriting in spatial and temporal properties, we were inspired by work on dynamic movement primitives (DMP) that are used for modeling attractor behaviors of autonomous nonlinear dynamical systems with the help of statistical learning techniques \cite{Ijspeert:2013bt}. We will not make use of the DMPs ability to learn and reproduce trajectories, but will configure a damped spring system similarly to a DMP and instead of applying a forcing term $f$ that activates the system's nonlinear dynamics over time we make use of an obstacle avoidance technique mentioned in \cite{Hoffmann:kk} which we adpoted and inverted its force to actually move towards the goal in a goal forcing function $g$ (see eq. \ref{eq:dampened_spring_system_sup}). The reason for this is that when we simply applied the spring system to each goal sequentially, we would accelerate toward and slow down at the goal. To keep up the momentum we need to look ahead several goals $x_{i+3}$ (here 3 steps ahead) in the core spring system, but with a goal forcing function that sequentially tries to visit each goal $x_i$. $\alpha, \beta, \gamma$ and $\mu$ are constants that specify the behavior of the system. $\varphi$ is the angle to the goal (or its velocity) and $y$ is the current position.
\begin{equation}
\begin{split}
\varphi &= \varphi_{x_i} - \varphi_{\dot{y}} \\
\dot{\varphi} &= \gamma ~\varphi ~e^{-\mu |\varphi|} \\
g &= (x_i - y) ~\dot{\varphi}  \\
\ddot{y} &= \alpha (\beta (x_{i+3} - y) - \dot{y}) + g
\label{eq:dampened_spring_system_sup}
\end{split}
\end{equation}
The resulting acceleration $\ddot{y}$ will be twice integrated before it is applied as an environmental state space position $\vartheta(t) = y$.

Once the goal is reached the Motor level $L_M$ will send a direct signal to the Vision level $L_V$, where the current location will be evaluated, given the goal location that was also received by $L_V$.

% \vskip 0.2in
% \begin{thebibliography}{1}
% \bibitem{Hoffmann:kk} Hoffmann, H., Pastor, P., Park, D.-H., \& Schaal, S. (2009). Biologically-inspired dynamical systems for movement generation: Automatic real-time goal adaptation and obstacle avoidance (pp. 2587–2592). Presented at the 2009 IEEE International Conference on Robotics and Automation (ICRA), IEEE. \url{http://doi.org/10.1109/ROBOT.2009.5152423}
% \bibitem{Ijspeert:2013bt} Ijspeert, A. J., Nakanishi, J., Hoffmann, H., Pastor, P., \& Schaal, S. (2013). Dynamical movement primitives: learning attractor models for motor behaviors. MIT Press
% , 25(2), \url{328–373. http://doi.org/10.1162/NECO_a_00393}
% \bibitem{Mateos:2017bp} Mateos, D. M., Riveaud, L. E., \& Lamberti, P. W. (2017). Detecting dynamical changes in time series by using the Jensen Shannon divergence. Chaos: an Interdisciplinary Journal of Nonlinear Science, 27(8), 083118–14. \url{http://doi.org/10.1063/1.4999613}
% \end{thebibliography}

\end{document}